\documentclass{article}
\usepackage{graphicx} % Required for inserting images
\usepackage{longtable}
\usepackage{amssymb}
\usepackage{comment}
\usepackage{amsmath}
\usepackage{authblk}
\usepackage{hyperref}

\title{A Quantitative Approach to Estimating Bias, Favouritism and Distortion in Scientific Journalism}
\author[1]{Raghavendra Koushik}
\author[2,3]{Hector Zenil\thanks{Corresponding author: hector.zenil@kcl.ac.uk}}

\affil[1]{\normalsize\text{ }Oxford Immune Algorithmics, Oxford University Innovation and London Institute of Healthcare Engineering, Oxford and London, U.K.}

\affil[2]{\normalsize\text{ }Department of Biomedical Computing, School of Biomedical Engineering and Imaging Sciences, Faculty of Medicine and Life Sciences \& King's Institute for AI, King's College London, U.K.}

\affil[3]{\normalsize\text{ }Cancer Interest Research Group, The Francis Crick Institute, London, U.K.}

\date{}

\begin{document}

\maketitle

\begin{abstract}
While traditionally not considered part of the scientific method, science communication is increasingly playing a pivotal role in shaping scientific practice. Researchers are now frequently compelled to publicise their findings in response to institutional impact metrics and competitive grant environments. This shift underscores the growing influence of media narratives on both scientific priorities and public perception. In a current trend of personality-driven reporting, we examine patterns in science communication that may indicate biases of different types, towards topics and researchers. We focused and applied our methodology to a corpus of media coverage from three of the most prominent scientific media outlets: Wired, Quanta, and The New Scientist--spanning the past 5 to 10 years. By mapping linguistic patterns, citation flows, and topical convergence, our objective was to quantify the dimensions and degree of bias that influence the credibility of scientific journalism. In doing so, we seek to illuminate the systemic features that shape science communication today and to interrogate their broader implications for epistemic integrity and public accountability in science. We present our results with anonymised journalist names but conclude that personality-driven media coverage distorts science and the practice of science flattening rather than expanding scientific coverage perception.\\

\end{abstract}

%OUTLETS: SEE WHERE THE PROF FROM SCOTLAND PUBLISHES HIS PAPERS ON SCIENCE REPRODUCIBILITY. TAKE THE MUTUAL INFORMATION OF EACH PAPER PER AUTHOR FOR A RANKING OF REUSE OF SAME SOURCES

\section{Introduction}

Although there is a great deal of research on bias in scientific research~\cite{HeesenBright2025,ZverevaKozlov2021}, there is little research on scientific communication bias. Some of the research explains bias due to the sales pressure of publishers on journalists~\cite{Dempster2022,Wuehrl2024} some other shows how the dynamics of information dissemination in digital media can be modelled as a complex adaptive system in which manipulative content, echo-chambers and polarised opinion clusters emerge from structural and dynamical network effects, a perspective that helps us understand how systematic bias can propagate through news ecosystems~\cite{Caldarelli2025}.

Science communication has become a key mechanism to cultivate and maintain public trust in science. Science journalists are therefore gatekeepers of what and how the public reads and learns science and what research is given coverage or not. Given that much scientific research is publicly funded, the accuracy, characterisation, and transparency of how science is communicated should, however, also be subject to accountability or disclosure, akin to standards upheld in professional journalism in other areas, including politics. This is particularly critical in an era of increasing political polarisation and epistemic fragmentation, where scientific discourse is often entangled in broader societal divides and scientific communication is no longer stranger to such politisation and polarisation, from climate change to scientific practice with now even scientific influencers online in audio and video social media channels.

Communication of scientific knowledge has traditionally been regarded an adjunct to the scientific method--important but largely external to the processes of hypothesis formation, experimentation, and validation. However, in the current research landscape, this view is no longer tenable. Increasingly, science communication is shaping the trajectory of scientific inquiry itself. Researchers are encouraged and often required to publicise their work to demonstrate social impact, meet funding requirements, and ensure visibility within an intensely competitive academic environment. Funding bodies, particularly in the United Kingdom and across Europe, have embedded impact and outreach metrics into grant assessment frameworks, effectively institutionalising the imperative to communicate. As a result, the boundaries between scientific content, public perception, and media strategy have become increasingly blurred.

In recent years, science communication has undergone a profound transformation, driven by the rise of personality-led dissemination and the dominance of social media. A new class of science communicators has emerged, ranging from media-savvy researchers to independent influencers, often operating without institutional oversight or methodological transparency. These individuals now wield disproportionate influence over which scientific outputs gain visibility, favour, or neglect, frequently guided by personal interest, popularity metrics, or monetisation strategies. Scientific writers, in particular, have positioned themselves as powerful arbiters of attention, often insulated from scrutiny except for editorial checks aligned with commercial imperatives. This unaccountable authority introduces new forms of structural bias with the potential to distort both scientific legitimacy and public trust.

This entanglement raises critical concerns regarding the credibility and accountability of science communication. The assumption that scientific results speak for themselves is challenged by the realities of media framing, selective reporting, and audience targeting. Public trust in science—an essential element in democratic societies where research is largely publicly funded—is now mediated through complex networks of information exchange, many of which operate outside traditional scholarly ecosystems. In an era of political polarisation, misinformation, and declining institutional trust, the framing and transmission of scientific knowledge have consequences far beyond academic recognition. The credibility of science itself may be influenced, or even compromised, by the manner in which it is communicated.

Compounding this issue is the emergence of personality-driven and social media–mediated science communication. Researchers with large followings, journalists with editorial autonomy, and independent influencers now shape what scientific developments gain attention and which are ignored. This environment fosters the rise of new science communicators who operate largely without oversight, editorial review, or peer accountability. They may promote or dismiss scientific claims based on subjective interests, platform incentives, or alignment with audience preferences. The visibility of a scientific article is no longer solely determined by its rigour, novelty, or peer reception, but increasingly by its media appeal and endorsement by influential figures.

Scientific writers, in particular, are occupying a growing space of power within this communication ecosystem. Armed with narrative skill and editorial backing, they have become central nodes in the dissemination of scientific knowledge to the public. However, unlike scientists whose work is subject to peer review and public scrutiny, many of these communicators are responsible only for commercial editorial policies. The decision of what to cover, how to frame it, and whether to support or ignore particular scientific findings is often shaped by strategic, economic, or ideological considerations rather than by standards of epistemic integrity. This lack of accountability introduces a subtle but consequential layer of bias in the representation of science, a bias that can skew public understanding, influence policy debates, and even affect the direction of future research.

In light of these developments, there is a growing need for systematic and empirical tools to evaluate the credibility and bias of science communication.

Here, we analyse a curated corpus of science coverage from leading popular science outlets, including Wired, Quanta Magazine, and New Scientist, that span a period of five to ten years. Our approach captures the relational and temporal dynamics of media narratives: how topics cluster, how certain figures or institutions dominate coverage, and how influence propagates through citation, repetition, and linguistic resonance.

By quantifying these patterns, we aim to identify markers of structural and editorial bias that can undermine the credibility of science communication. We argue that such bias not only affects how science is perceived by the public but may also feedback into the scientific process itself, prioritising certain research directions over others based on visibility rather than intrinsic value. Ultimately, our goal is to contribute to the development of accountability frameworks for science communication: tools that can assess its integrity, track its influence, and help ensure that the dissemination of scientific knowledge upholds the values of transparency, fairness, and public service on which science itself depends.

\section{Methods}

We compiled a corpus of science-related articles published between 2014 and 2024 from three major English-language outlets: \textit{Wired}, \textit{New Scientist}, and \textit{Quanta Magazine}. For each outlet, the corpus was processed in several stages to extract, filter, and analyse person-level mention data.  

The first step involved the aggregation of mentions. Each article contained a \texttt{mention\_counts} JSON dictionary, from which we extracted the frequency of named individuals. These counts were then summed across all articles belonging to a given outlet, producing an aggregated record of the total number of times each person was mentioned.  

A filtering stage was then applied to exclude certain categories of names. Specifically, we removed any name that exactly matched an entry on a blacklist consisting of three groups: (i) deceased or non-living scientists, such as Albert Einstein, Isaac Newton, Marie Curie, and Stephen Hawking; (ii) non-scientist public figures and celebrities, including Donald Trump, Elon Musk, and Joe Biden; and (iii) spurious single-token entries, including stray punctuation marks or isolated words that did not correspond to personal names. This ensured that the dataset focused exclusively on living scientists and avoided contamination by irrelevant or misleading entries.  
To maintain consistency across all analyses, no occupational or role-based classification was performed; names are treated as they appear in article text, and minor credit prefixes (e.g., ``By'') or role descriptors were only cleaned when explicit.

To protect journalists' names, an anonymisation procedure was introduced. The complete set of author names for each outlet was extracted from the JSON data, and each name was assigned to a unique anonymised identifier of the form \texttt{Author\_XXX}, where the marker \texttt{XXX} represented a three-digit index. Whenever an aggregated mention corresponded to one of these authors, their real name was replaced by the corresponding anonymised identifier. Mentions of other individuals who had passed the blacklist and name format filters were retained in their original form.  

Following filtering and anonymisation, the data were represented as dictionaries mapping each person to their total number of mentions. Then these mention counts were sorted in descending order to produce ranks of individuals by prominence within each outlet. To assess the degree of inequality in attention, we calculated the Gini coefficient for each distribution. The coefficient was computed using the standard formula:  

\[
  G = \frac{2 \sum_{i=1}^n i\,x_{(i)}}{n \sum_{i=1}^n x_{(i)}} - \frac{n+1}{n},
\]

where \(x_{(i)}\) denotes the ordered mention counts and \(n\) the number of individuals. This measure ranges from 0, indicating perfectly even representation, to 1, indicating extreme concentration of attention.  

In log-log histograms, cumulative distribution functions and Zipf rank ---frequency plots, we excluded the top 1\% of mention counts to facilitate the reading and avoid logarithmic artifacts. These extreme outliers reflected exceptionally high repetition by a small subset of authors and were removed only for the purposes of visual clarity. All quantitative analyses, including percentile statistics, Gini coefficients, and top-ten rankings, were performed on the complete datasets. The log-log histograms displayed the frequency distribution of mention counts, while the Zipf plots presented the log-transformed relationship between rank and frequency.  

Two further measures were computed to capture overall editorial focus. The first was the total number of mentions in all living scientists, which reflects the cumulative editorial weight devoted to scientific figures in each outlet. The second was the number of distinct scientists mentioned at least once, which provides an indication of the breadth of coverage.  

Finally, to examine differences in long-tail editorial behaviour, we computed the reverse cumulative distribution function (1-CDF) of mention counts. This measure shows the proportion of scientists who were mentioned at least \(x\) times, thus highlighting the persistence of repeated exposure. For plotting purposes, the top 0.5\% of values in each outlet were trimmed to prevent distortion by rare but extreme outliers. Both axes of these plots were presented on logarithmic scales, emphasising differences in editorial treatment across the long tail of mention distributions.

\subsection{Topic and Sentiment Analysis}

Beyond raw mention counts, we also examined the linguistic framing of articles by analysing the titles of all science-related articles in our data set.

\paragraph{Topic Frequency.}
We tokenised and vectorised all article titles per outlet using \texttt{ Scikit-learn}'s \texttt{CountVectorizer} (excluding English stopwords and tokens shorter than three characters). For each outlet, we produced frequency counts of all terms and plotted the top 20 most common words. This reveals dominant thematic emphases (e.g., ``quantum,'' ``AI,'' ``cosmos'') that characterise the editorial framing of scientific stories.

\paragraph{Sentiment Analysis.}

We quantified the sentiment polarity of the titles using the \texttt{VADER} sentiment analyser from NLTK, which computes a compound score between $-1$ (highly negative) and $+1$ (highly positive). Each article title was scored, and the sentiment distributions at the outlet level were visualised using boxplots. This allowed us to compare whether certain outlets systematically frame scientific developments in more optimistic, neutral, or negative tones.

\section{Results}

The person-author network revealed highly skewed patterns. Some scientists were mentioned by particular journalists with significantly higher frequency, suggesting recurring narrative focus or bias.

Key findings:
\begin{itemize}
\item \textbf{Repetitive Focus:} Certain authors mentioned the same individual (e.g. ``Lee Cronin”) in more than five different articles, with cumulative mentions exceeding 15 that we had identified and reported before despite or as a result of the controversial nature of his group results.
\item \textbf{Topic-Specific Bias:} Certain scientists dominated specific topic categories (e.g., quantum physics or reproducibility) across multiple articles.
\item \textbf{Asymmetrical Representation:} A small number of journalists repeatedly promoted a small cluster of scientists, while most others were mentioned once or not at all.
\end{itemize}

\begin{figure}[hbp!]
  \centering
  \includegraphics[width=0.85\textwidth]{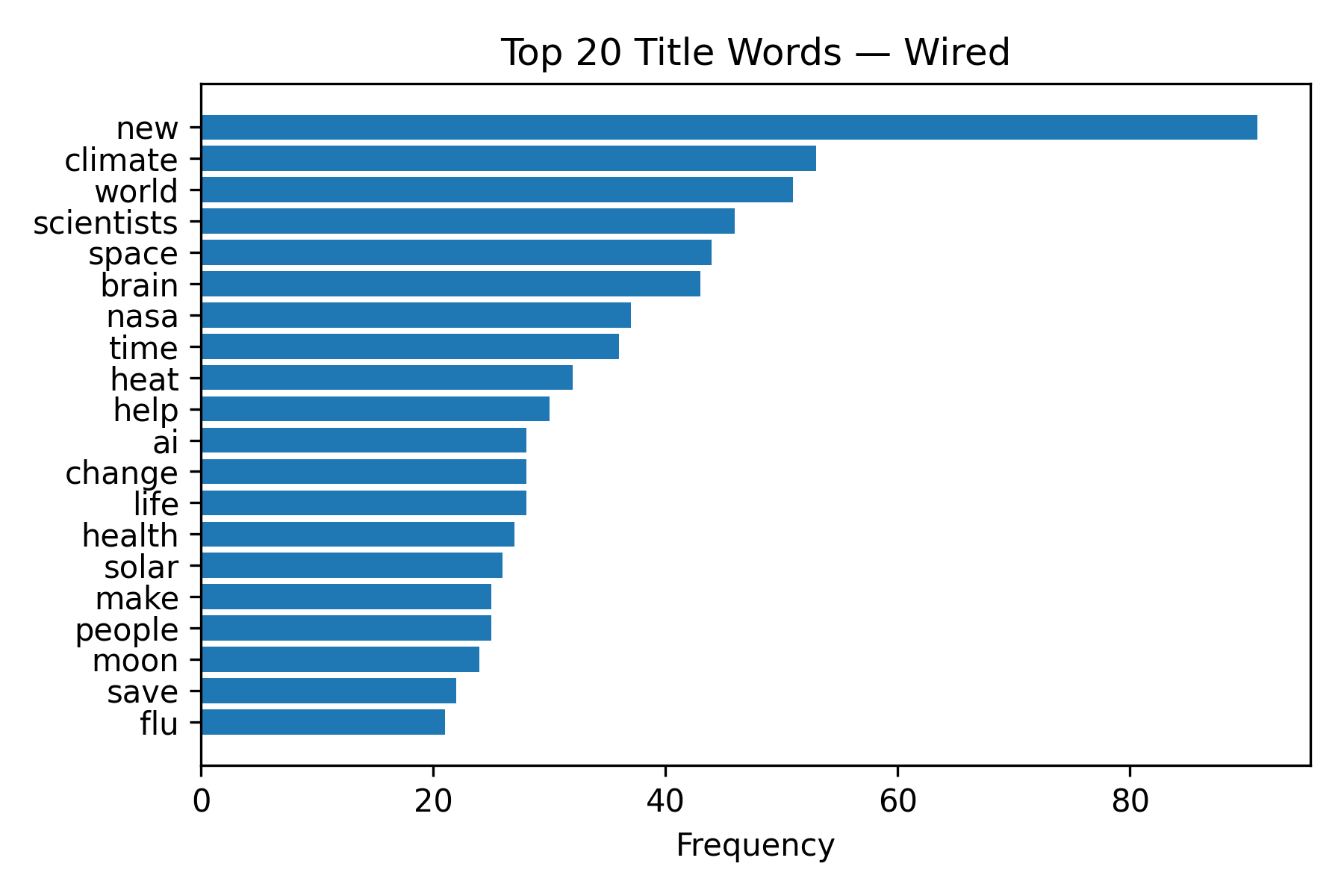}
  \caption{\textbf{Top 20 Title Words in Wired.} Frequent use of terms such as ``AI,'' ``quantum,'' and ``data'' indicates emphasis on technology-driven narratives.}
  \label{fig:topics_wired}
\end{figure}

\begin{figure}[hbp!]
  \centering
  \includegraphics[width=0.83\textwidth]{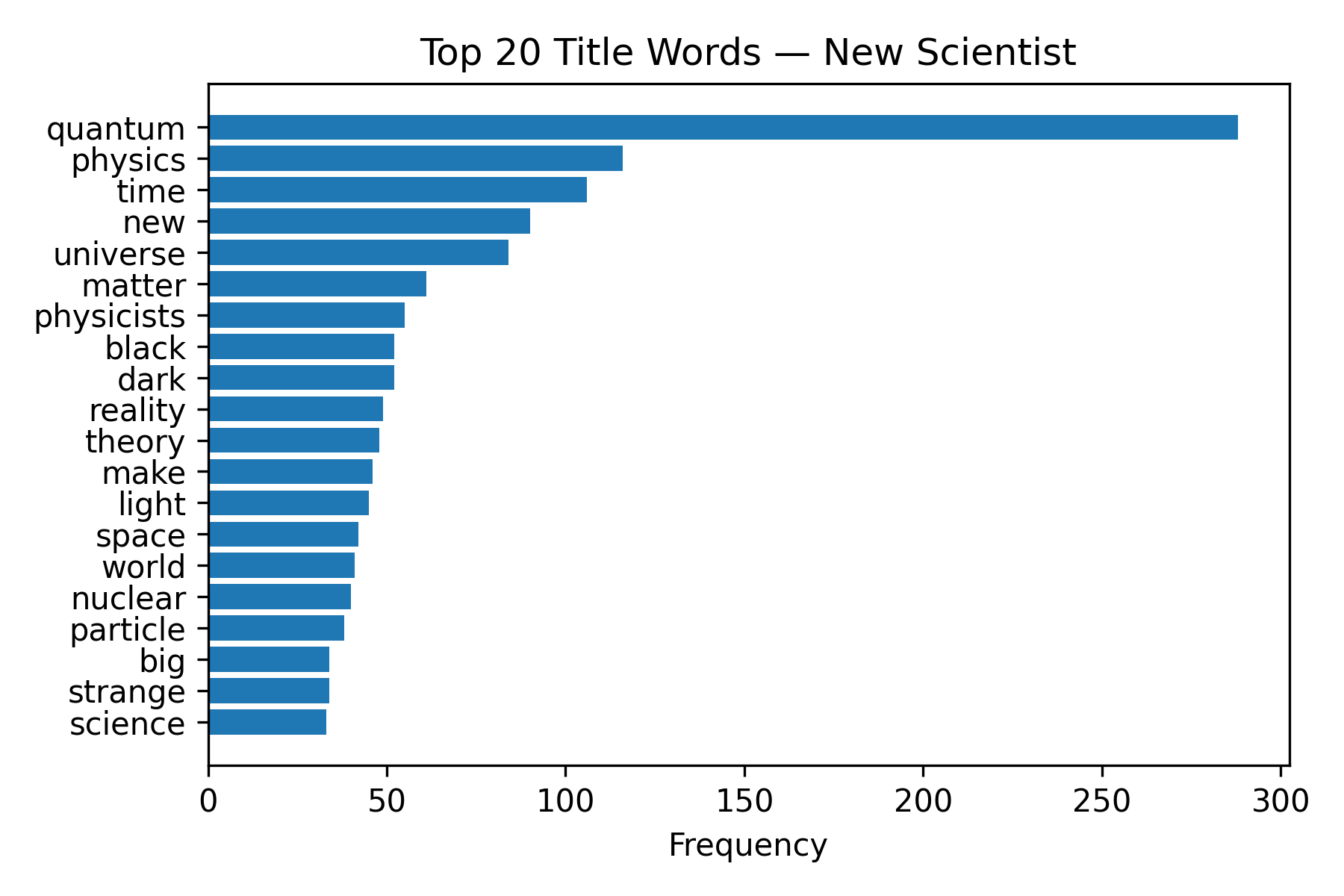}
  \caption{\textbf{Top 20 Title Words in New Scientist.} Coverage is more broadly distributed across cosmology, health, and climate-related terms.}
  \label{fig:topics_ns}
\end{figure}

\begin{figure}[hbp!]
  \centering
  \includegraphics[width=0.87\textwidth]{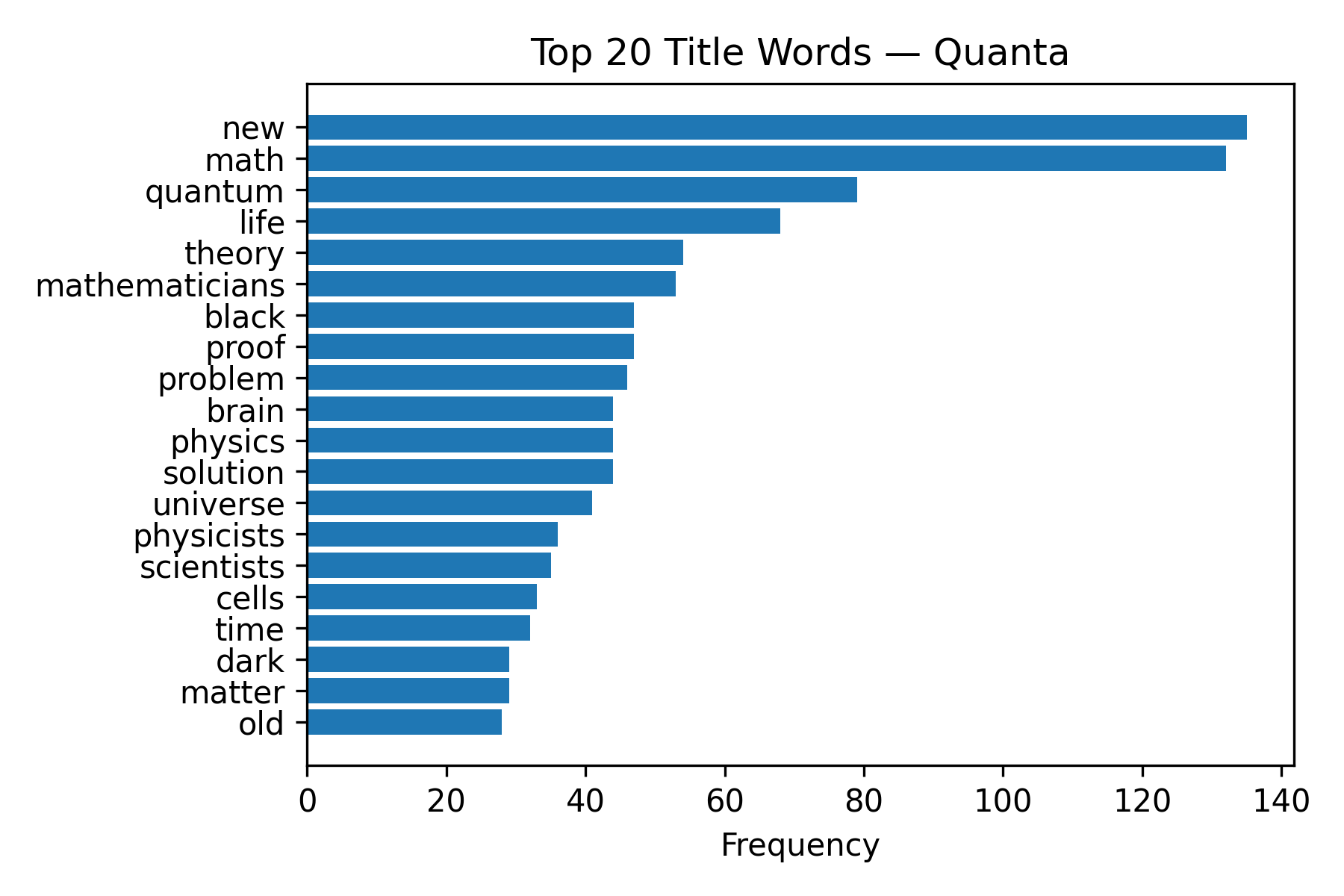}
  \caption{\textbf{Top 20 Title Words in Quanta.} Quanta emphasizes mathematical and foundational science topics, with terms like ``theory,'' ``proof,'' and ``quantum'' dominating.}
  \label{fig:topics_quanta}
\end{figure}

\begin{figure}[hbp!]
  \centering
  \includegraphics[width=0.85\textwidth]{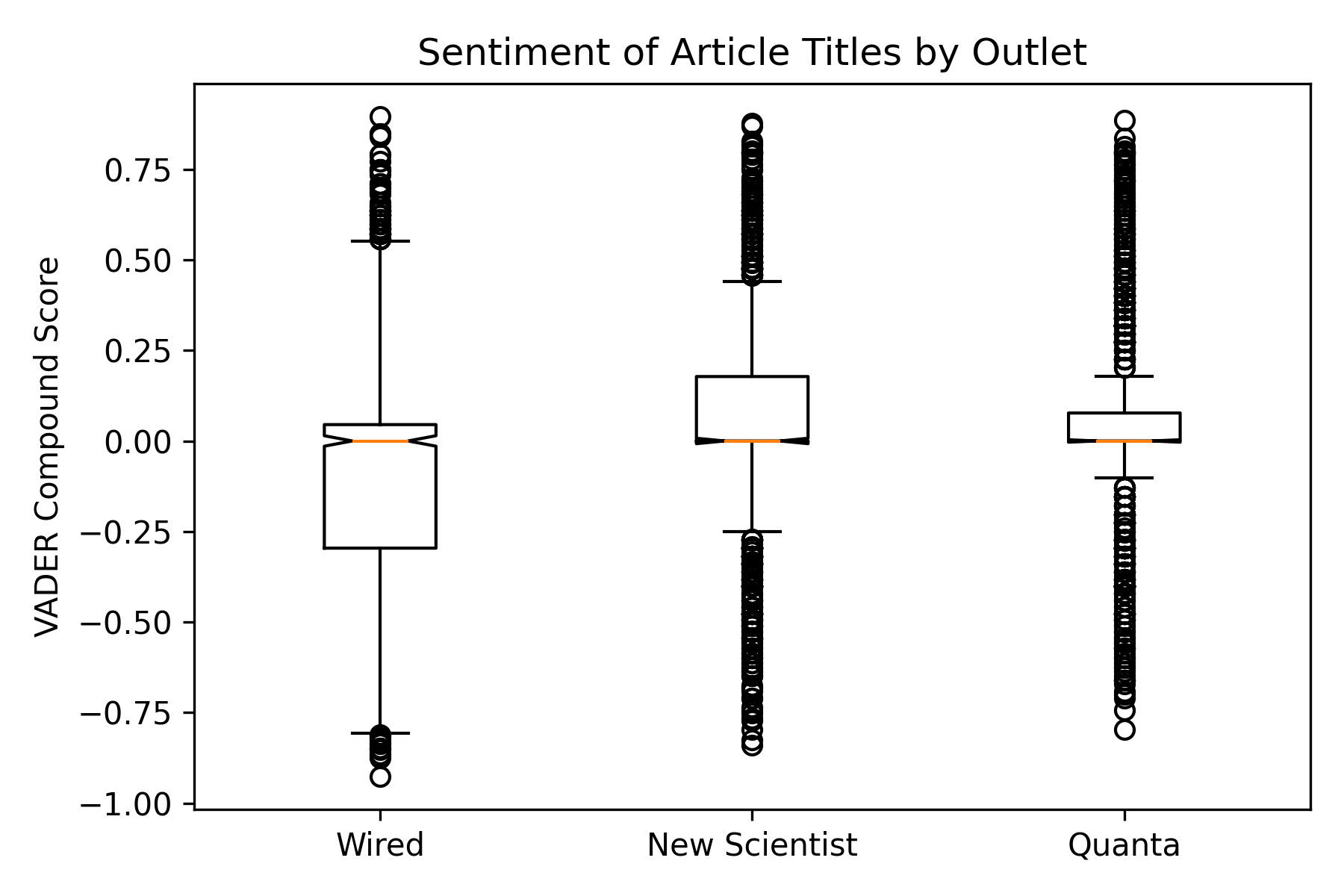}
  \caption{\textbf{Sentiment of Article Titles by Outlet.} VADER compound scores show Quanta and New Scientist framing titles in a more positive tone, while Wired displays more variability and a wider sentiment spread suggesting a more balanced to critical approach. The higher score, the more positive, while the more lower score the more negative.}
  \label{fig:sentiment_titles}
\end{figure}

\begin{comment}

\begin{center}
\renewcommand{\arraystretch}{1.2}
\begin{longtable}{p{0.4\textwidth} r}
\caption{Top Most Mentioned Individuals per Author (Quanta) after removal of `universal' authors that are no longer academically active as they are death and are universally recognised (Turing, Einstein, Newton, Feynman, G\"odel, Hawking, Schr\"odinger, Er\"os, Maxwell, Born, and Fermat).}\label{balltable}\\
\endfirsthead

\textbf{Person} & \textbf{Mentions} \\
\endhead

\input{quanta_top_mentions_table_cleaned.tex}

\end{longtable}
balltable
\end{center}
\end{comment}

All results were presented using anonymised author identifiers (e.g. 'Author A'), with visualisations that include:
\begin{itemize}
\item Interactive bar charts of mentions per scientist per author
\item Co-mention networks highlighting narrative cliques
\item Tabular summaries of top-mentioned individuals per author
\end{itemize}

% Section: Top Repetition Bias / Abuser Index

\subsection{Repetition Bias and Scientific Distortion}

To evaluate potential favouritism or narrative bias in scientific journalism, we computed a \textbf{Repetition Bias Index} for each journalist from each outlet. This index aggregates how often an author mentions the same living individual across multiple articles. High values may indicate overexposure or preference for specific scientists. The results indicate a strong skew. Although most authors exhibit wide diversity in the scientists they mention, a few show disproportionately high repetition for certain individuals. For example, Author A mentions \textit{Scientist X} in 15 separate articles, often repeating the name multiple times per piece. As a manual validation example, we tracked all mentions of certain authors in articles authored by one of the Top 20 writers for each outlet to validate our name normalisation and repetition tracking strategy. The reported concentration of repetitions, when not widespread in other authors, may signal editorial bias, favoritism, or narrative reinforcement around specific figures.

\begin{figure}[h!]
\centering
\includegraphics[width=\textwidth]{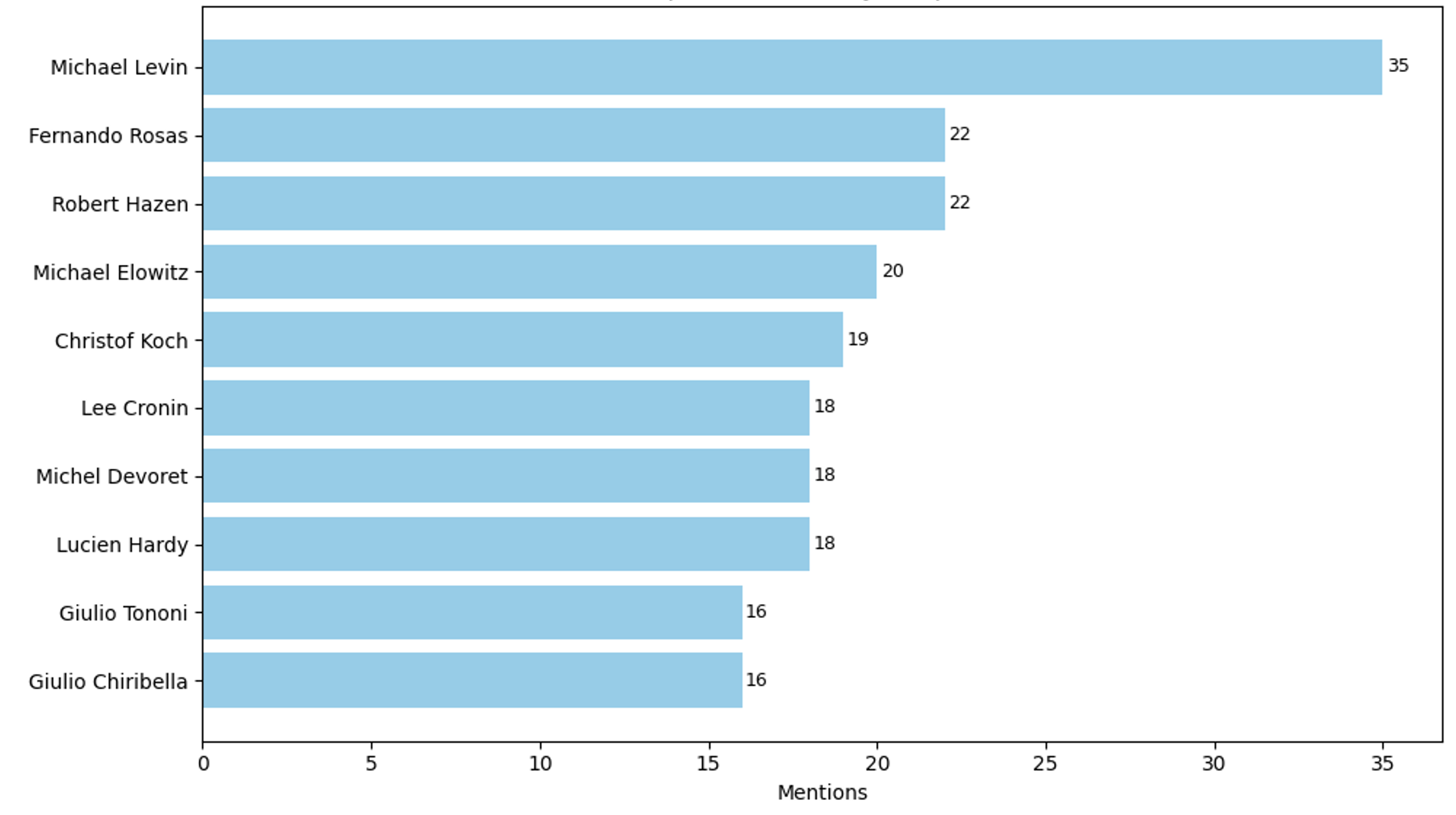}
\caption{Example of individual mention bias from a top 20 scientific anonymised scientific journalist from Quanta. Recent Wired practice of publishing verbatim versions of Quanta articles may be driven Wired's bias towards Quanta. 
}\label{fig:cdf}
\end{figure}

While the bias reported in Fig.~\ref{fig:cdf} can be explained by the interest of the author in a selection of topics, the selection of topics seems biased and potentially unbalanced across topics and authors. Even taking a selection of topics of choice, each area certainly has more authors than this selection. Among this selection, there are authors such as Christof Koch, Lee Cronin, and Giulio Tononi who have recently participated in controversies and disputes highly cited by the scientific media, potentially fuelling the controversy, as this manifesto suggested~\cite{iit1,iit2}. This seems to favour and reward controversial authors that are unlikely to be also the most scientifically accurate or relevant, just as happens in regular media where there is an asymmetry between sources of fake news and fact checkers.

In the case of Tononi and Koch, for example, their theory (or a hypothesis) of consciousness, or Integrated Information Theory (IIT), has been the object of controversy and large numbers of peers' public disapproval~\cite{iit1,iit2} criticising both the theory authors for allowing overexposure and misleading statements in the scientific media and the media for their frenzy-ness towards controversy and this particular theory, giving the impression that IIT had a much greater weight over other hypotheses or was the mainstream view of the field as having `solved' consciousness and evidence was in their favour. Regardless of the value of the theory/hypothesis, the discomfort that the bias provoked and the scientific media role they played, was reported as a disservice to the field.

%\subsection*{Case Study: Wired Magazine}
%We also conducted a detailed analysis of \textit{Wired}'s science section to explore potential narrative bias and repetition of person mentions.

All journalist identities were anonymised, deceased or historical scientists were filtered out, and public figures such as Elon Musk, Donald Trump, Joe Biden, etc. removed.

\begin{figure}[hbp!]
  \centering
  \includegraphics[width=0.5\textwidth]{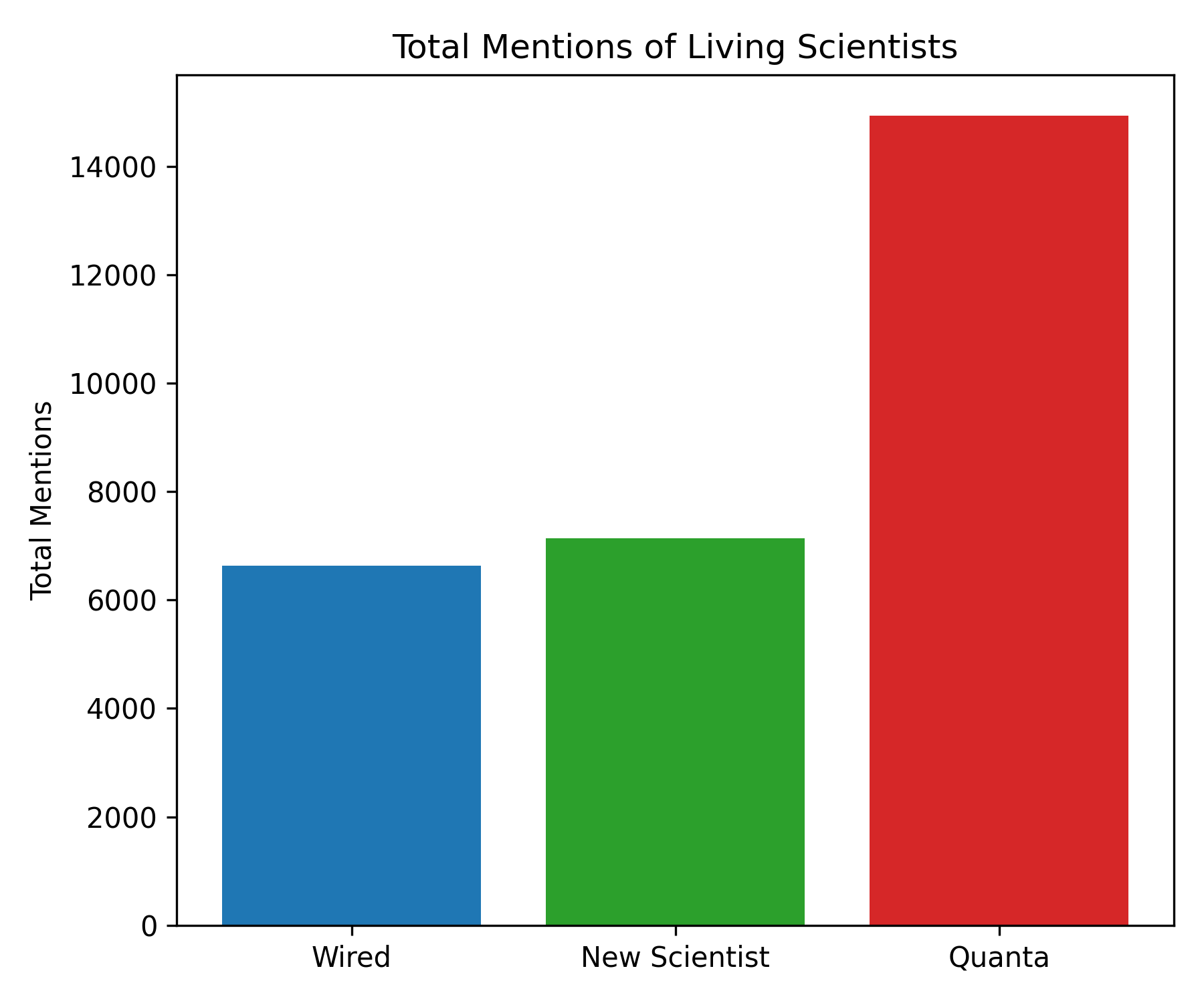}\includegraphics[width=0.5\textwidth]{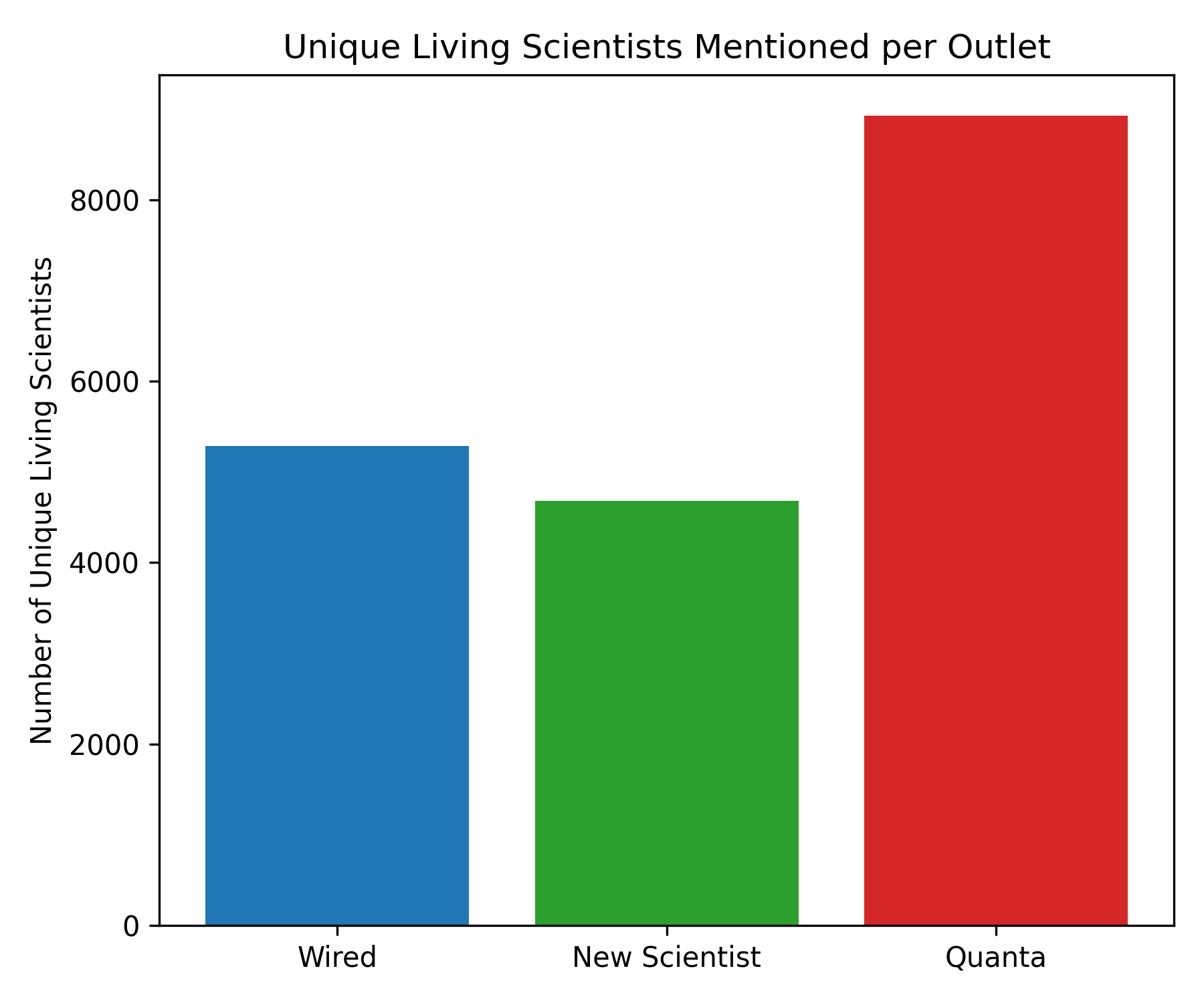}
  \caption{Left: Quanta features more than double the total number of living scientist mentions compared to Wired and New Scientist, indicating a significantly stronger editorial focus on highlighting scientists.  Right: Quanta mentions a significantly larger number of distinct living scientists, indicating not only depth but also editorial breadth in science coverage.}
  \label{fig:total_mentions}
\end{figure}

To further quantify mention concentration, we generated:

% \item \textbf{Cumulative Distribution Function (CDF):} The CDF plot (Figure~\ref{fig:cdf}) reveals that the top 200 scientists account for roughly 40% of mentions, and the top 1000 cover only about 55%, indicating a steep skew toward a small group.
\begin{itemize}
    
\item A reverse CDF plot (Fig.~\ref{fig:total_mentions} left) showing the proportion of living scientists who are mentioned at least $x$ times per outlet. After trimming the top 0.5\% of extreme outliers to improve clarity, we observe that \textit{Quanta} consistently features a higher proportion of scientists with repeated mentions. This highlights a stronger editorial emphasis on a more broader and deeper engagement with living scientists compared to \textit{Wired} and \textit{New Scientist}.

\item A mention-count histogram (Fig.~\ref{fig:total_mentions} right) that shows over 5200 scientists with fewer than 10 mentions each, and a long tail of a few with more than 50 mentions, underscoring heavy imbalance.
\end{itemize}

\begin{comment}
    
\begin{figure}[h!]
\centering
\includegraphics[width=0.8\textwidth]{figures/wired_cdf.png}
\caption{Cumulative distribution of scientist mentions in Wired.}
\label{fig:cdf}
\end{figure}
\begin{figure}[h!]
\centering
\includegraphics[width=0.8\textwidth]{figures/wired_histogram.png}
\caption{Histogram of scientist mention counts in Wired.}
\label{fig:hist}
\end{figure}

\end{comment}

\begin{figure}[hbp!]
  \centering
  \includegraphics[width=0.55\textwidth]{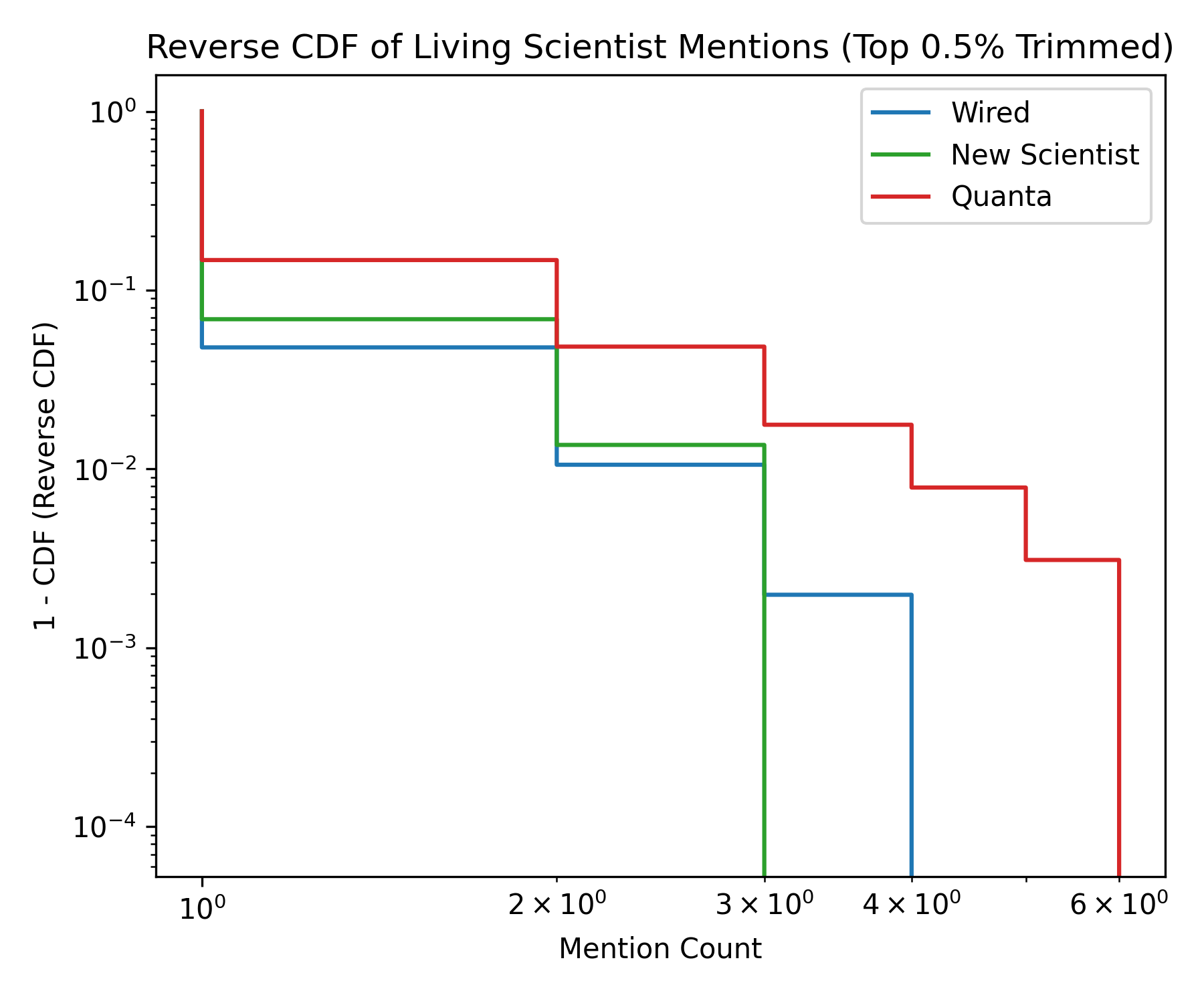}\includegraphics[width=0.55\textwidth]{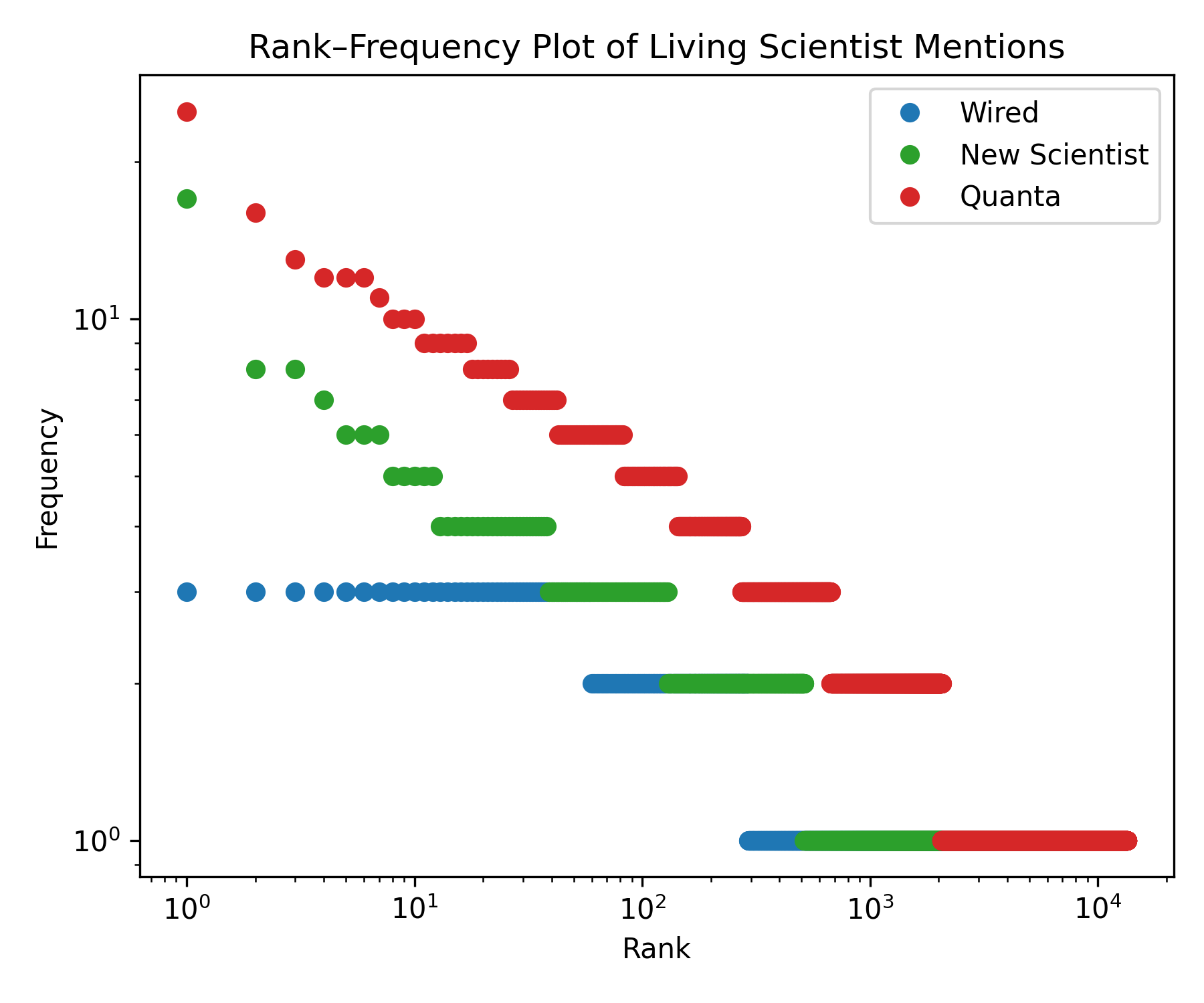}
  \caption{Left: Reverse Cumulative Distribution Function of Scientist Mentions. Right: Rank–Frequency Plot of Scientist Mentions. Both plots show a strong bias of Quanta to report personality-based results and over representation of the same researchers in their articles followed by Wired and New Scientist.}
  \label{fig:reverse_cdf}
\end{figure}

The results indicate skewness of $\approx 42.6$ and Zipf slope of $\approx -0.35$, confirming that even after rigorous filtering, coverage remains dominated by a small core of scientists.

%\subsection*{Comparison and Implications}
%When these metrics are applied to Quanta Magazine and The New Scientist, we will directly compare the skewness values, Zipf slopes, and CDF thresholds to determine which outlet exhibits the highest or lowest concentration of voices. These comparisons will serve as quantitative indicators of structural bias.

%\subsection{Top 10 Most-Mentioned Living Scientists}

\begin{table}[h!]
\centering
\resizebox{\textwidth}{!}{%
\begin{tabular}{c l r | c l r | c l r}
\multicolumn{3}{c}{\textbf{Wired}} &
\multicolumn{3}{c}{\textbf{New Scientist}} &
\multicolumn{3}{c}{\textbf{Quanta}} \\ \hline
\# & Scientist & \% & \# & Scientist & \% & \# & Scientist & \% \\ \hline
1 & Celine Halioua & 28.57 & 1 & Chanda Prescod-Weinstein & 33.12 & 1 & Steven Strogatz & 19.00 \\
2 & Gül Dölen & 18.01 & 2 & Carlo Rovelli & 16.72 & 2 & David Kaplan & 14.78 \\
3 & Andrew Hessel & 16.15 & 3 & Roger Penrose & 11.99 & 3 & Peter Greenwood & 12.40 \\
4 & Sáenz Romero & 6.83 & 4 & Geraint Lewis & 7.57 & 4 & Patrick Honner & 10.29 \\
5 & Andrew Wakefield & 6.21 & 5 & Sean Carroll & 7.26 & 5 & Roger Penrose & 9.23 \\
6 & Ben Tabib & 6.21 & 6 & Vlatko Vedral & 5.99 & 6 & Terence Tao & 8.18 \\
7 & John Scott & 5.59 & 7 & Frank Wilczek & 4.73 & 7 & Petr Stepanek & 6.86 \\
8 & David Marchant & 4.35 & 8 & Matt Strassler & 4.42 & 8 & Yunger Halpern & 6.60 \\
9 & David Lochridge & 4.35 & 9 & Juan Maldacena & 4.10 & 9 & Nima Arkani-Hamed & 6.60 \\
10 & Vitale Brovarone & 3.73 & 10 & Lee Smolin & 4.10 & 10 & Glen Palmer & 6.07 \\
\hline
\end{tabular}%
}
\caption{Top 10 most‐mentioned \emph{living scientists} per magazine, shown as percentages of the total mentions among each outlet's top ten scientists. Percentages were computed using the same filtering logic as the Gini and histogram analyses (full‐name entities only, excluding deceased and non‐scientist figures, and verified through Wikipedia classification).}
\end{table}

%\subsection{Inequality Metrics (Gini Coefficients)}

\begin{table}[h!]
\centering
\begin{tabular}{lc}
\hline
\textbf{Magazine} & \textbf{Gini Coefficient} \\
\hline
Quanta Magazine  & 0.18                     \\
Wired            & 0.08                     \\
New Scientist    & 0.08                     \\
\hline
\end{tabular}
\caption{Gini coefficients of mention distributions, computed over living scientists only showing Quanta's personality-driven approach to scientific journalism.}
\end{table}

%\subsection{Comparative Visualisations}

\begin{comment}
    
\begin{figure}[hbp!]
  \centering
  \includegraphics[width=0.7\textwidth]{figures/histogram.png}
  \caption{Log–Log Histogram of Scientist Mention Counts (Wired, New Scientist, Quanta).}
  \label{fig:combined_histogram}
\end{figure}

\end{comment}

\section{Conclusions and Limitations}

A methodological limitation of this study is that a lot of manual curation was needed given the nature of most outlets to be behind paywalls with strong anti-web scraping policies even for research purposes. This would suggest that outlets should probably do this exercise themselves and probably release it publicly to promote transparency. Another limitation of this study is that it is biased itself towards a very small number of outlets. Although these outlets can be said to have some global reach especially in English-speaking countries, they are by no means representative of all outlets in the world. However, the same methodology can be applied in a larger study. What this study shows is that it can stratify and find differences that seem to mirror the style of each outlet and can provide some accountability feedback to help these outlets regulate themselves.

As for the results, the bar plots shown in Fig.~\ref{fig:total_mentions} highlight Quanta's strong emphasis on covering living individuals, often focussing on the person rather than the work. Quanta accounts for more total mentions than Wired and New Scientist combined. Although all outlets displayed this bias, this is more pronounced in Quanta and may create a greater distortion of the importance of certain people or work and may lack the balance and neutrality expected from scientific journalism.

In distribution-based plots (Fig.~\ref{fig:reverse_cdf}), Quanta consistently dominates the high-mention tail, meaning that not only are more scientists covered, but many receive repeated attention. The reverse CDF plot, in particular, shows that a higher proportion of scientists are mentioned at least 2 to 5 times more in Quanta relative to the other outlets.

\begin{enumerate}

  \item \textbf{New Scientist (Gini = 0.08)} has the highest even mention distribution, with a relatively balanced coverage between scientists.
  \item \textbf{Wired (Gini = 0.08)} is moderately skewed, with some emphasis on recurring figures.
  \item \textbf{Quanta (Gini = 0.18)} shows the numerically most concentrated distribution, reflecting repeated mentions of a selective group of scientists.
\end{enumerate}

While bias in journalism may arise from personal preferences, ideological inclinations, or professional specialisation of a journalist, professionalism lies in the ability to overcome such influences. The true test of journalistic integrity is not the elimination of perspective, which is unattainable, but the conscious effort to recognise and moderate one's own predispositions in the pursuit of balanced reporting. This disciplined detachment is not only an ethical obligation, but also a methodological safeguard, ensuring that journalism continually strives to be a reliable source of accurate and impartial information. Consequently, journalist's professionalism should be measured by how effectively they can distance themselves from their personal choices to defend neutrality and fairness.

We have reached a point where the media and scientific journalists decide what and how science is consumed by the public and some outlets focus more on
topics than on personalities, our analysis shows that both topic and researcher selection are more pronounced in personality-driven outlets. Among the three analysed, Quanta exhibited the strongest bias, followed by Wired.

From hypotheses such as Integrated Information Theory and Assembly Theory that have generated a large amount of controversy in the community~\cite{Jaeger,at1,at2,at3}, the media's attention in controversial theories, even if largely rejected by the communities in the way they are covered\cite{iit1,blog}, seems to be sending the wrong signal by amplifying attention.

Identifying and reporting such patterns is a necessary first step towards improving accountability in science communication and promoting fairer coverage across disciplines, rather than reinforcing a Matthew effect in which the most visible figures attract even greater attention. The results reported here suggest that certain individuals are systematically favoured by outlets and writers and that personality-driven is a greater distortion factor.

\section*{Data and Code Availability}
All processed data and analysis code supporting this study are available at:  
\url{https://github.com/zenil-colab/bias-in-science-communication}

\end{document}